\documentclass[pra,showpacs,showkeys,nofootinbib]{revtex4}

\usepackage{amsmath,amssymb}
\usepackage{epsfig}
\usepackage{color}

\newcommand{\n}{\nonumber}
\newcommand{\be}{\begin{equation}}
\newcommand{\ee}{\end{equation}}
\newcommand{\bea}{\begin{eqnarray}}
\newcommand{\eea}{\end{eqnarray}}

\newcommand{\fDs}{~_sD^\beta_x}

\begin{document}

\title{\textbf{Similarity solutions for a class of  Fractional Reaction-Diffusion equation }}
\author{Choon-Lin Ho}
\affiliation{Department of Physics, Tamkang University
Tamsui 25137, Taiwan}



\begin{abstract}

This work studies exact solvability of a class of fractional reaction-diffusion equation with the Riemann-Liouville fractional derivatives on the half-line in terms of the similarity solutions.  We derived the conditions for the equation to possess scaling symmetry even with the fractional derivatives. Relations among the scaling exponents are determined, and the  appropriate similarity variable introduced.  With the similarity variable we reduced the differential partial differential equation to a fractional ordinary differential equation. Exactly solvable systems are then identified by matching the resulted ordinary differential equation with the known exactly solvable fractional ones.  Several examples involving the three-parameter Mittag-Leffler function (Kilbas-Saigo function) are presented.  The models  discussed here turn out to correspond to superdiffusive systems. 

\end{abstract}

\pacs{05.40.Fb, 02.50.Ey, 05.10.Gg}

\keywords{Fractional calculus, Reaction-Diffusion equation, spacetime-dependent
diffusion and reaction, similarity method}

\maketitle


\section{Introduction}

The well-known Brownian motion is characterized by a mean-squared displacement relation $\langle x^2 \rangle\sim t$ between displacement $x$ and time $t$.  However, there are situations in which  $\langle x^2 \rangle\sim t^\gamma(\gamma\neq 1)$ is not linear in time.  Such are called the anomalous diffusions -- superdiffusive for $\gamma>1$, subdiffusive for $\gamma<1$.
Anomalous diffusions have been found in many physical situations.   For examples, charge carrier transport in amorphous semiconductors, nuclear magnetic resonance diffusometry in percolative, and porous systems, Rouse or reptation dynamics in polymeric systems, transport on fractal geometries, and many others (see  \cite{MK} for a nice review).

   The Brownian motion is Markovian in nature.  This means each new step in the motion depends only on the present state, and is independent of the the previous states.  Anomalous diffusion arises, on the contrary, from some memory effect of previous states, or as a result of some fractal structure of the background space, or due to some non-linear interaction inherent in the system, etc.  Accordingly, the most commonly employed approaches 
  to model anomalous diffusion are by fractional differential equations for diffusion equations \cite{D1,D2,D3,D4,D5,D6,D7,D8,D9,D10,D11,D12}, Fokker-Planck equations \cite{FP1,FP2,FP3,FP4,FP5,FP6,FP7,FP8,FP9}, and reaction-diffusion equations \cite{FRD1,FRD2,FRD3,FRD4,FRD5,FRD6,FRD7}; by various nonlinear extensions of the Fokker-Planck equations \cite{N1,N2,N3,N4,N5}; or both \cite{N6} .
   
   Among various ways to model anomalous diffusion, the approach based on differential equations with fractional differential operators \cite{B1,B2,KST,GKMR}  has attracted much attention in recent years.  This is mainly because  these equations, despite the presence of fractional derivatives, can still be linear (except the nonlinear type), and as such many familiar techniques in ordinary calculus can still be employed.  Other than statistical physics, fractional differential equations have also found applications recently in many other areas, such as quantum mechanics \cite{Q1,Q2,Q3}, fluid mechanics \cite{F1,F2,F3,F4,F5,F6}, and nonlinear science \cite{NS1,NS2,NS3,NS4,NS5,NS6,NS7}.
  
Several models of anomalous diffusions, mostly with different definitions of time-fractional derivatives, have been proposed and studied in the past. 
In this work we would like to consider exact solvability of the following class of spatial fractional Reaction-Diffusion equations (FRDE)  (with the Fokker-Planck equation as its  special case),
\be
\frac{\partial }{\partial t}P(x,t)= \fDs\left(D(x,t)P(x,t)\right) + R(P,x,t),
\label{FRD1}
\ee
where $P(x,t)$ is the particle distribution function,
$D(P,x,t) $ is the diffusion coefficient, and $R(P, x,t)$ the
reaction term  which accounts for the  local reaction or external force.
This equation describes the anomalous change of concentration/population of a substance/species distributed in space  under the influence of two processes: local reaction which modify the concentration/population, and diffusion which causes the substances/species  to spread in space. 
We use the term  ``particle" to denote generally the number of basic member of a substance or a species.
Our interest in spatial fractional  derivative is motivated by the fact that diffusion equations with spatial fractional derivatives allow one to account for certain superdiffusive processes, called the L\'evy flights, which have divergent mean squared displacement $\langle x^2 \rangle\sim t$ \cite{D3,D4,D5,D6,D7}.  Here we also include possible effect of the external reaction .

As several definitions of fractional derivatives are available, we shall adopt the more commonly discussed Riemann-Liouville type of fractional derivative defined by
\bea
\fDs f(x) &=&\frac{d^n}{dx^n} ~_sI^{n-\beta}_x\,f(x),\\
~_sI^\alpha_x\, f(x) &=&\frac{1}{\Gamma(\alpha)}\int^x_s (x-x^\prime)^{\alpha-1} f(x^\prime)\, dx^\prime, ~~ n=[\beta]+1.\n
\label{RL}
\eea
$[x]$ is the integral part of$x$.
The domain we shall consider in this paper is mainly the half line $x\in [0,\infty)$. Unlike most previous works on fractional reaction-diffusion equations, we leave the possibility that $D$ and $R$ could be functions of $P$.

   As with differential equations based on the usual derivatives, exact solutions of fractional differential equations are hard to obtain.  The most commonly employed methods to find solutions of such equations are by the integral transform methods, and eigenfunctions expansions \cite{MK}. Very often, if not always, such methods involve separation of variables.

 In this work, we shall be concerned with solving the FRD (\ref{FRD1}) 
 in terms of the similarity solutions. This is motivated by our recent works on similarity solutions of the ordinary differential equations \cite{ho1,ho2,ho3,ho4,ho5,ho6}. One advantage of the similarity method is that it allows one to
reduce the partial differential equation under consideration to an
ordinary differential equation which is generally easier to solve, provided that the original equation
possesses proper scaling property under certain scaling
transformation of the basic variables.  This method has been applied to study specific models of diffusion/reaction equations with fractional time-derivative \cite{D5,D8,FRD5} and fractional space-time-derivartive \cite{D6}, and in fractional diffusion-wave equation \cite{D7}.  Here we would like to extend our previous consideration to the fractional case.  Our work differs from the previous ones in that we leave the coefficients of the diffusion and reaction terms unfixed, and determine their form from the requirement of scale invariance.   

The  plan of  this paper  is as follows:
Sect.~II discusses
the conditions for the FRDE to possess scaling symmetry.
Sect.~III introduces the corresponding similarity variable and scaling forms of the relevant functions, which are used to
 reduce the FRDE into an ordinary fractional differential
equation.  The equation of continuity is discussed in Sect.~IV which helps to identify two types of  scaling behaviours of the FRDE.
Some examples of exactly solvable FRDE  are discussed in Sect.~V to VII. 
Sect.~VIII concludes the paper.

\section{Scaling of Reaction-Diffusion equation}

If Eq.\,(\ref{FRD1}) possesses certain
scaling symmetry, then it admits similarity solutions. Consider the scale transformation
\be
x=\epsilon^a \bar{x}\;\;\;,\;\;\; t=\epsilon^b \bar{t},
\ee
where the scale factor $\epsilon$ and the two scaling exponents $a$ and $b$ are real parameters. Suppose
under this transformation, the distribution function, the diffusion coefficient and
the reaction term  scale as
\bea
P(x,t) &=& \epsilon^c \bar{P}(\bar{x},\bar{t}),\n
\\
D(P,x,t)  &=&\epsilon^d \bar{D}(\bar{P}, \bar{x},\bar{t}),
\\
R(P,x,t) &=&\epsilon^e \bar{R}(\bar{P}, \bar{x},\bar{t}),\n
\label{scale}
\eea
where the scaling exponents $c$, $d$ and $e$ are some real parameters.  Then 
\bea
\frac{\partial }{\partial t}P(x,t) &=&  \epsilon^{c-b}\frac{\partial \bar{P}}{\partial
\bar{t}},\\
 \fDs(\left(D(x,t)P(x,t)\right)  &=&
 \epsilon^{c+d} \fDs\left(\bar{D}(\bar{x},\bar{t}) \bar{P}(\bar{x},\bar{t})\right) \n\\
&=&\epsilon^{-\beta a+c+d}~ _{\bar s}D^\beta_{\bar x}\left(\bar{D}(\bar{x},\bar{t}) \bar{P}(\bar{x},\bar{t})\right),\n\\
 {\bar s} &=&\epsilon^{-a}s.\n
\eea
Here we have used the identity
\be
\fDs f(\bar{x}=bx)=b^\beta~_{(bs)}D^\beta_{\bar x} f(\bar{x}),
\ee
which can be easily checked from Eq.(\ref{RL}).

Written in
the transformed variables, Eq.(\ref{FRD1}) becomes
\be\
\epsilon^{c-b}\frac{\partial \bar{P}}{\partial
\bar{t}}=\epsilon^{-\beta a+c+d}~_{\bar s}D^\beta_{\bar x}\left(\bar{D} \bar{P}\right) + \epsilon^e\, \bar{R}.
\label{FRD2}
\ee
For simplicity and clarity of presentation, here and below we shall often omit the independent variables in a function.

Eq.(\ref{FRD2}) has the same functional form as Eq.(\ref{FRD1}), 
if and only if  the scaling indices satisfy $b=\beta a-d=c-e$, and ${\bar s}=\epsilon^{-a}s=s$. The latter condition implies that 
 $s$ can only be $s=0$ (for the half real line) or $s=-\infty$ (for the whole real line).  In this work, we shall only consider $s=0$ as it is in this case that several exact solutions are available \cite{KST,GKMR}.

\section{Similarity variable and scaling forms}

When Eq.\,(\ref{FRD1}) possess scaling symmetry, it can be solved by the similarity method. 
 Similarity method allows one to transform the
 partial differential equation to an ordinary differential equation  through some new
independent variable (called similarity variables), which are
certain combination of the old independent variables such that
they are scaling invariant, i.e., no appearance of parameter
$\epsilon$, as a scaling transformation is performed.

The similarity variable $z$ is defined by
\be
z\equiv\frac{x}{t^{\alpha}}, ~~\mbox{where}~
\alpha=\frac{a}{b}\;\;\;,\; b\neq 0\;.
\ee
Eq.\,(\ref{scale}) is realized if we assume the following scaling forms of the distribution function, the diffusion and the reaction terms in terms of $z$:
\be
P(x,t)=t^\mu y(z), ~~ D(P,x,t)=t^\nu \rho(z), ~~ R(P,x,t)=t^\lambda \sigma(z).
\ee
From Eq.\,(\ref{scale}) together with the scaling conditions $b=\beta a-d=c-e$,  one has
\be
\mu=\frac{c}{b}, ~~ \nu=\frac{d}{b}=\beta\alpha-1, ~~\lambda=\frac{e}{b}=\mu-1.
\ee
Thus $\alpha$ and $\mu$ are the only two independent scaling exponents of the RDE.

In terms of these scaling forms, Eq.~(\ref{FRD1}) reduces to an ordinary fractional differential equation
\be
_{\bar s}D^\beta_z\left(\rho y\right) + \alpha z\frac{dy}{dz}-\mu y+ \sigma(z)=0.
\label{OFRD}
\ee
Note that when $\mu=-\alpha$, which we will encounter below, Eq.\,(\ref{OFRD}) reduces to
\be
_{\bar s}D^\beta_z\left(\rho y\right) +\frac{d}{dz}\left( \alpha z y\right)+ \sigma(z)=0.
\label{ODE_e}
\ee

The results discussed thus far can be employed in two ways. 

First, given a specific fractional reaction-diffusion system of interest in applied sciences or engineering, one can use the procedures described here to check if this model possesses scaling symmetry, and if it does, reduce it to an fractional ODE (\ref{OFRD}) by introducing the appropriate similarity variable, and then solve the resulted ODE by some appropriate methods, eg., analytic, asymptotic, approximate, and numerical. 

Second, by matching the Eq.\,(\ref{OFRD})  with an ODE with appropriate domain, initial conditions and boundary condition, one can determine the corresponding D, and R functions, and then construct a fractional reaction-diffusion system that can be solved as long as the ODE can be solved by any method. Here we will  concentrate on exactly solvable ODEs.

Before we proceed to discussing  some exactly solvable systems, let us first consider the conditions imposed by
the continuity in the change of the particle number of the system,
i.e., the equation of continuity.

\section{Equation of continuity}

The total number $\cal N$ of the system is related to the density function $P(x,t)$ by
\be
{\cal N}=\int_{\cal D}\,P(x,t)\,dx=t^{\alpha+\mu}\int_{\cal D}\,
y(z)\,dz,
\label{N}
\ee
where $\cal D$ is the domain of the independent variable.  For simplicity, we use the same notation $\cal{D}$ for both the variable $x$, and the corresponding similarity variable $z$. 

 Eq.\,(\ref{N}) distinguishes two different situations: $\alpha+\mu\neq 0$ and $\alpha+\mu=0$: $N$ is conserved if and only if  $\mu=-\alpha$.

The time rate of change of $\cal N$ is
\be
\frac{d{\cal N}}{dt}=(\alpha+\mu)t^{\alpha+\mu-1} \left(\int_{\cal D}\,
y(z)\,dz\right).
\label{dN}
\ee
But from Eq.\,(\ref{FRD1}) one has
\bea
&& ~~~~~~~ \frac{d{\cal N}}{dt}=\int_{\cal D}\,\frac{\partial P}{\partial t}\,dx \\
&=& \int_{\cal D}\,\fDs(\left(D(x,t)P(x,t)\right)  + \int_{\cal D}\, R(P,x,t)\,dx.\n
\eea

In terms of the similarity variable $z$, 
\bea
&&~~~~~~~~  (\mu+\alpha) t^{\alpha+\mu-1} \int_{\cal D}\,y\, dz \n\\
=
&&t^{\alpha+\mu-1}\int_{\cal D}\,_{\bar s}D^\beta_z\left(\rho y\right) \,dz  + t^{\alpha+\lambda}\int_{\cal D}\, \sigma(z)\,dz.
\eea
In view of $\lambda=\mu-1$, one has
\be
(\mu+\alpha)  \int_{\cal D}\,y\, dz=
\int_{\cal D}\,_{\bar s}D^\beta_z\left(\rho y\right) \,dz  + \int_{\cal D}\, \sigma(z)\,dz.
\ee
With  the identity
\be
_{\bar s}D^\beta_z\left(\rho y\right)  = \frac{d}{dz} ~_{\bar s}D^{\beta-1}_z\left(\rho y\right),
\label{d-1}
\ee
which follows from the definition (\ref{RL}), we have 
\be
 (\alpha+\mu)\int_{\cal D}\,
y(z)\,dz= \int_{\cal D}\, \sigma(z)\,dz + \Delta \left[~_{\bar s}D^{\beta-1}_z\left(\rho y\right)\right]_{\partial {\cal D}}.
\label{EOCa}
 \ee
 Here $\partial{\cal D}$ denotes the boundaries of the domain $\cal D$, and $\Delta[\cdots]_{\partial {\cal D}}$ the difference of the terms in the bracket at the boundaries.  But one can rewrite Eq.\,(\ref{OFRD}) as 
 \be
\frac{d}{dz} \left[~_{\bar s}D^{\beta-1}_z\left(\rho y\right)+\alpha z y\right]-(\mu + \alpha)y-\sigma(z) =0.
\label{OFRD2}
\ee
 By integrating  Eq.(\ref{OFRD2}) and comparing the resulted equation with (\ref{EOCa}), we see that the equation of continuity is equivalent to
 \be
\Delta[ z y]_{\partial {\cal D}}=0.
\label{EOCa2}
\ee
This is the same as that obtained for the ordinary convection-diffusion-reaction equation, which includes the Fokker-Planck equation and the reaction-diffusion equation as sub-cases \cite{ho6}.
 
 It should be noted that in cases where the number of particle $\cal N$ cannot be defined, i.e., if the integral in Eq.\,(\ref{N}) is divergent, then Eq.\,(\ref{EOCa})  or (\ref{EOCa2}) are not satisfied. Two such examples are presented in \cite{ho5}.

In what follows, we shall present some exactly-solvable examples for $\mu=-\alpha$ and $\mu\neq -\alpha$.

\section{Cases with $\mu= -\alpha$ and $\sigma=0$}

The case with $\mu= -\alpha$ and $\sigma=0$ corresponds to fractional diffusion.  In this case
Eq.\,(\ref{OFRD2}) gives
\be
_{\bar s}D^{\beta-1}_z\left(\rho y\right)  +\alpha z y= {\rm constant}.
\label{FRD_e}
\ee
In this work we shall only consider cases with  the constant equals zero, i.e.,
\be
_{\bar s}D^{\beta-1}_z\left(\rho y\right) +\alpha z y=0.
\label{FRD_e0}
\ee

As mentioned in Sect. III,  we will only be interested in obtaining exactly solvable reaction-diffusion system by matching Eq.\,(\ref{OFRD}) with an exactly solvable ODE. Unfortunately,  there are not many known exact fractional ODEs in the literature. So we shall present examples based on the exact fractional ODEs available from \cite{KST}.

Consider a special class of problems on the half-line $x\geq 0$, so $z\geq 0$ and $s=\bar s=0$,  with
$\rho(z)=z^{-q}$ ($q$ a real constant). Let $~Y(z)=\rho\, y(z)$, or $y(z)=z^q\, Y(z)$. 
Eq.\,(\ref{FRD_e0}) becomes
\be
_0D^{\beta-1}_z Y(z)  +\alpha z^{q+1} Y(z) =0.
\label{FRD_Y}
\ee
The diffusion coefficient is $D(x,t)=t^{-\nu} z^{-q} = t^{1-\beta\alpha} z^{-q}$.  The order $\beta$ of the fractional differential operator,  the scaling exponent $\alpha$, and $q$ are the three independent parameters in this system.

According to Theorem 4.13 in Section 4.2.6 of \cite{KST} (summarized in Appendix C),  the general solution of 
(\ref{FRD_e0}) is a linear combination of the following solultons
\bea
Y_j(z)&=& z^{(\beta-1)-j}\,E_{\beta-1, 1+\frac{q+1}{\beta-1}, 1+\frac{q+1-j}{\beta-1}}\left(-\alpha z^{\beta+q}\right), \n
\\
  j &=&1,2,\ldots, n, ~n=-[-(\beta-1)],~~\beta>1,
\label{Yj}
\eea
with the boundary condition (\ref{EOCa2}) 
\be
\Delta[z^{q+1} Y(z)]_{\partial {\cal D}}=0.
\ee
Here $E_{\alpha,m,l}(z)$ is the three-parameter  Mittag-Leffler function (or the Kilbas-Saigo function) (Appendix A).
The general solution of Eq.\,(\ref{FRD1}) is
$P(x,t)=t^{-\alpha} \sum_j c_j  y_j(z), y_j(z)=z^q\, Y_j(z)$.
For normalized $P(x,t)$ one must have $\beta+q-n-1\ge 0$, or $q\geq n+1-\beta$.

Fig.\,1 depicts graphs of $y_j (z)$ for $1<\beta\leq 2$ (with $n=1, j=1$ and $q=n+1-\beta=2-\beta$) , and $2\leq \beta <3$ (with $n=2, j=1,2$ and $q=n+1-\beta=3-\beta$).
The numerical results indicate that  for $\beta>2$, $y_j(z)$ could be negative for some values of $z$. As $P(x,t)\geq 0$, being a distribution function, the fractional diffusion equation discussed here only makes sense for $\beta\leq 2$.  So $n=j=1$, and
\be
P(x,t)=N \frac{x^{\beta+q-2}}{t^{\alpha(\beta+q-1)}} \times 
E_{\beta-1, 1+\frac{q+1}{\beta-1}, 1+\frac{q}{\beta-1}}\left(-\alpha z^{\beta+q}\right),
\label{P1}
\ee
where  $N$ is the normalization constant.

Normalized $P(x,t)$ requires $\beta+q-2\ge 0$, or $q\geq 2-\beta$.
There are two possible situations : 1) if $q=2-\beta$, then $P(x,t)>0$ at $x=0$; 2) if $q> 2-\beta$, then $P(x,t)=0$ at $x=0$.

The diffusion coefficient in the standard diffusion equation for the Brownian motion is a constant.  Without loss of generality we set $D(x,t)=1$.  This implies  $\rho=1$ and $\nu=\beta\alpha-1=0$. So $\alpha$ and $\beta$ are tied by $\alpha=1/\beta$. The fractional diffusion equation reduces to the standard diffusion equation for $\beta=2$ ($n=1$) and $q=0$.  In this case Eq.\,(\ref{P1}) becomes, according to Eqs.(B1), (B4) and (B6),
\bea
P(x,t)&\propto \frac{1}{\sqrt{t}} E_{1,2,1}(-\alpha z^2)\n\\
         &\propto \frac{1}{\sqrt{t}} E_{1,1}(-\frac{\alpha}{2}z^2)\\
         &\propto \frac{1}{\sqrt{t}} e^{-\frac{\alpha x^2}{2t}},\n
\eea
which is the well-known solution of the diffusion equation.

We will present some examples with $0<\beta\leq 2$ and $\alpha=1/\beta$. 

In Figs.\,2 and 3 we show the evolution of the distribution function $P(x,t)$  with $q=2-\beta$ and $q> 2-\beta$, respectively.
The former case can be considered as describing the deformed Brownian motion  (on the half-line) by $\beta$. 
It is seen that at large times $t$, $P(x,t)$  for $\beta<2$ is slightly larger than that for $\beta=2$ (the standard Brownian). So the fractional diffusion equation discussed here can be considered as a model for superdiffusive system. 

We note here that a type of superdiffusive motion,  so-called the L\'evy flight, has been modeled by a diffusion equation with the Riesz-Weyl type fractional space-derivative \cite{D3,D4}, which is different from that employed here.

For $\beta=1$, Eq.\,(\ref{P1}) is not valid.  However, in this case Eq.\,(\ref{FRD_e}) can be solved easily. 
Now we have
\be
_0D^0_z\left(\rho y\right)  + \alpha z y=  C, ~~C={\rm constant},
\ee
which is the reduced fractional ODE for the equation
\be
\frac{\partial }{\partial t}P(x,t)= \frac{\partial }{\partial x} \left(D(x,t)P(x,t)\right).
\ee
As $_0D^0_z (\rho y ) =\rho y$, the solution is
\be
y(z)=\frac{C}{\rho(z)+\alpha z}.
\label{y_b1}
\ee
For the choice $\alpha=1/\beta=1,~ C=2/\pi$, and $\rho(z)=1-z+z^2$, we have the Cauchy distribution (on the half-line) 
\be
y(z)=\frac{2}{\pi\left(1+z^2\right)}.
\ee
The probability density $P(x,t)=t^{-1} y(z)$ turns out to correspond to the distribution function of the L\'evy flight described by the fractional diffusion equation based on the Riesz-Weyl fractional space-derivative with constant diffusion coefficient  \cite{D3,D4}.

We note here that for some choices of $\rho(z)$, the solution $y(z)$ in (\ref{y_b1}) may describe systems in which the number of particle $\cal N$ is not defined, and hence the boundary condition (\ref{EOCa2}) is not valid.  For example,  if we take $\rho(z)=1$.

\section{Cases with $\mu= -\alpha$ and $\sigma\neq 0$}

Eq.\,(\ref{ODE_e}) is not easy to solve exactly in general when $\sigma(z)\neq 0$.  But  we can find some exactly solvable systems
for $\sigma(z)$ a gradient of some function $\tau(z)$, i.e., $\sigma(z)=-d\tau(z)/dz$.  Under this situation, Eq.\,(\ref{ODE_e}) becomes
\be
\frac{d}{dz} \left[_{\bar s}D^{\beta-1}_z\left(\rho y\right) + \alpha z y-\tau\right]=0,
\ee
or
\be
_{\bar s}D^{\beta-1}_z\left(\rho y\right) + \alpha z y-\tau=0,
\label{FRD_e1}
\ee
where we have absorbed the constant of integration into the function $\tau(z)$.

\subsection*{$\bullet$ Fokker-Planck type}

If $\tau(z)$ is proportional to $y(z)$ and $z$, i.e., $\tau(z)=-\gamma z y(z)$ for some constant $\gamma$, the diffusion equation is
\be
_{\bar s}D^{\beta-1}_z\left(\rho y\right) + (\alpha + \gamma) z y=0,
\ee
which is just Eq.\,(\ref{FRD_e0}) with $\alpha$ replaced by $\alpha+\gamma$.  Hence solutions can be found as described in Sect. V with appropriate change of parameters.

\subsection*{$\bullet$ Non-Fokker-Planck type}

Let us consider solution of (\ref{FRD_e1}) with $s={\bar s}=0$, $y(z)=z^q\, Y(z), \rho(z)=z^{-q}$ and $\tau(z)=\sum_{r=1}^l f_r z^{\mu_r}$ ($f_r, \mu_r$ are real constants),
\be
_0D^{\beta-1}_z Y(z)  +\alpha z^{q+1} Y(z) =\sum_{r=1}^l  f_r z^{\mu_r} .
\label{FRD_e2}
\ee
According to Theorem 4.13 of \cite{KST} (Appendix C), for 
$n-1<\beta-1< n, q+1>-\{\beta-1\}$, and $\mu_r>-1 (r=1,\ldots, l)$
 the general solution of (\ref{FRD_e2}) is $Y(z)=Y_h(z) + Y_p(z)$, 
$Y_h(z)= \sum_{j=1}^n c_j Y_j(z)$ [$Y_j(z)$ given in (\ref{Yj})] is the solution of the homogeneous equation (\ref{FRD_e0}), and $Y_p(z)$ is a particular solution of (\ref{FRD_e2}) ($y_0(z)$ in Appendix C),
\bea
Y_p(z) &=& \sum_{r=1}^l \frac{\Gamma(\mu_r+1)\,f_r}{\Gamma(\mu_r+\beta)} z^{\beta-1+\mu_r} \n
\\
&&\times E_{\beta-1, 1+\frac{q+1}{\beta-1}, 1+\frac{q+1+\mu_r}{\beta-1}}\left(-\alpha z^{\beta+q}\right).
\eea
Then the $P(x,t)$ is 
\be
P(x,t)= N t^{-\alpha} z^q Y(z),
\label{P2}
\ee
where $N$ is the normalization constant.

In Fig.\,4 we present $P(x,t)$ for three different sets of $1<\beta\leq 2$ with $l=1$ and $\mu_r=-0.5$. Again it is seen that the for the choice of parameters, the smaller the value of $\beta$, the higher the value of $P(x,t)$ at large $x$ -- the reaction-diffusion system is superdiffusive.

\section{Cases with $\mu\neq -\alpha$}

From Eq.\,(\ref{dN}), $d\mathcal{N}/dt\neq 0$ if $\mu\neq -\alpha$, i.e., $\cal N$ need not be conserved. So in this case $P(x,t)$ is not interpreted as a probability density function, and so $\beta$ need not be restricted to be $\beta \leq 2$.  The reaction-diffusion equation (\ref{OFRD}) is in general difficult to solve.  Here we present a way to obtain a special class of exactly solvable system for (\ref{OFRD}).

Suppose
\be
\sigma(z)\equiv (\mu+\alpha)y-\frac{d}{dz}\tau(z),
\label{sigma}
\ee
for some function $\tau(z)$, then Eq.\,(\ref{OFRD2}) is integrated to be (with the integration constant absorbed into $\tau(z)$)
\be
_{\bar s}D^{\beta-1}_z\left(\rho y\right) + \alpha z y-\tau=0,
\ee
which is just Eq.\,(\ref{FRD_e1}).

Hence any choice of $\tau(z)$ that renders (\ref{FRD_e1}) exactly solvable defines an exactly solvable fractional reaction-diffusion system (\ref{OFRD}) with $\sigma(z)$ defined by (\ref{sigma}).  Particularly, the examples discussed in Sect. VI can be employed to define the corresponding systems in this case.

\section{Summary}

In this work we have studied exact solvability of a class of FRDE with the Riemann-Liouville fractional derivatives on the half-line in terms of the similarity solutions.  We have derived the conditions for the FRDE to possess scaling symmetry even with the fractional derivatives. Relations among the scaling exponents are determined, and the  appropriate similarity variable introduced.  With the similarity variable we reduced the partial differential equation to a fractional ordinary differential equation. Exactly solvable systems are then identified by matching the resulted  ODE  with known exactly solvable fractional ODEs.  Several examples were presented, which involve the  three-parameter Mittag-Leffler function (Kilbas-Saigo function). The FRDE's  discussed here turn out to correspond to superdiffusive diffusion. 

All the cases discussed in this work can be easily extended to FRDE's defined on the whole real line by simply changing $x$ to $|x|$, i.e., 
\be
\frac{\partial }{\partial t}P(x,t)= ~_sD^\beta_{|x|} \left(D(x,t)P(x,t)\right) + R(P, |x|,t).
\ee 
The function $P(x,t)$ in the negative half-line is just mirror image of those on the positive half-line (with the height reduced by a factor of two for normalization).

We note here that the procedure presented here can be adapted to fractional reaction-diffusion equations defined with other types of fractional derivative.

Also, it would be interesting to extend the present analysis to fractional versions of differential equations in other areas, such as  those in fluid mechanics, nonlinear science as mentioned in the Introduction, and mathematical biology \cite{E}.

\appendix

\section{Classical and generalized Mittag-Leffler Functions \cite{GKMR, KST}}

$\bullet$  Mittag-Leffler function

\be
E_\alpha(z)=\sum_{k=0}^\infty \frac{z^k}{\Gamma(\alpha k +1)}, ~~\alpha\in {\mathbb C}, ~~{\rm Re}\  \alpha >0.
\ee

$\bullet$ Two-parameter Mittag-Leffler function

\be
E_{\alpha, \beta}(z)=\sum_{k=0}^\infty \frac{z^k}{\Gamma(\alpha k +\beta)}, ~~\alpha, \beta \in {\mathbb C}, ~~{\rm Re}\  \alpha>0,
\ee

$\bullet$ Three-parameter Mittag-Leffler function (Prabhakar function) [see also \cite{Mai})

\be
E_{\alpha, \beta}^\gamma(z)=\sum_{k=0}^\infty \frac{(\gamma)_k\, z^k}{k!\,\Gamma(\alpha k +\beta)}, ~~\alpha, \beta, \gamma\in {\mathbb C}, {\rm Re}\  \alpha, \gamma>0,
\ee
where $(\gamma)_k=\gamma(\gamma+1)\cdots (\gamma+k-1)=\Gamma(\gamma+k)/\Gamma(\gamma)$ is the Pochhammer symbol.

\bigskip

$\bullet$ Three-parameter Mittag-Leffler function (Kilbas-Saigo function) [see also \cite{D12}]

\be
E_{\alpha, m,l}(z)=\sum_{k=0}^\infty c_k z^k,~~~~ z, l \in {\mathbb C},  {\rm}\ \alpha, m\in {\mathbb R}, ~~\alpha, m>0,
\ee
where
\be
c_0=1, ~~c_k=\prod_{i=1}^{k-1}\, \frac{\Gamma(\alpha[im+l]+1)}{\Gamma(\alpha[im+l+1]+1)}~~ (k=1,2,\ldots),
\ee
such that
\be
\alpha(j m+l) \neq -1,-2,\ldots (j=0,1,2,\ldots).
\ee


\section{Some identities \cite{GKMR}}

$\bullet$ $\alpha>0$
\be
E_{\alpha,1}(z)=E_\alpha(z).
\ee

$\bullet$ $ n\in {\mathbb N}, ~~\beta>0$
\be 
 E_{n,1,l}(z) =\Gamma(nl+1)\, E_{n,nl+1}(z).
\ee
\bigskip

$\bullet$ $m>0, l\in {\mathbb R}, jm+l\neq -1,-2, \dots. (j=0,1,2, \ldots)$
\be
E_{1,m,l}(z)=\Gamma\left(\frac{l+1}{m}\right) E_{1, \frac{l+1}{m}}\left(\frac{z}{m}\right).
\ee

\bigskip

$\bullet$ $l\neq -1,-2,\ldots$
\be
E_{1,1,l}(z)=\Gamma(l+1) E_{1,l+1}(z).
\ee

\bigskip

$\bullet$ $l= 0,1,2,\ldots$
\be
E_{1,1,l}(z)=l!\, E_{1,l+1}(z).
\ee

Particularly, 
\be
E_{1,1,0}(z)=E_1(z)=e^z.
\ee


\section{Cauchy problem for Inhomogeneous fractional differential equations with a quasi-polynomial free term \cite{KST}}

Here we summarize  Theorem 4.13 in Section 4.2.6 of \cite{KST}.

Consider the Cauchy problem for the Inhomogeneous differential equation of fractional order $\alpha>0$ with a quasi-polynomial free term
\bea
_{a^+}D^\alpha_x\,y(x)&=&\lambda(x-a)^\beta y(x) + \sum_{r=1}^l f_r (x-a)^{\mu_r} ,\n
\\ && ~~~~~~~~~~~~~  (a<x<b\leq \infty),\\
_{a^+}D^{\alpha-k}_x\,y(a^+)&=&b_k, ~~k=1,2,\ldots, n\equiv -[-\alpha],
\eea
where $\lambda, \beta, f_r, \mu_r\, (r=1,2,\ldots,l), b_k (k=1,\ldots,n) \in  {\mathbb R}$ are given real constants.

Let $n-1<\alpha < n, \beta>-\{\alpha\}$, and $\mu_r>-1 (r=1,\ldots, l)$. Then the Cauchy problem (B.1) and (B.2) has a unique solution in the space of functions locally integrable on $(a,b)$
\bea
y(x)&=&\sum_{j=1}^n \frac{b_j}{\Gamma(1+\alpha-j)} (x-a)^{\alpha-j}
\\
&&\times E_{\alpha, 1+\beta/\alpha, 1+(\beta-j)/\alpha}\left(\lambda (x-a)^{\alpha+\beta}\right) + y_0(x),\n
\eea
 where 
 \bea
 y_0(x) &=& \sum_{r=1}^l \frac{\Gamma(\mu_r+1)\,f_r}{\Gamma(\mu_r+\alpha+1)} (x-a)^{\alpha+\mu_r}
 \\
&&\times E_{\alpha, 1+\beta/\alpha, 1+(\beta+\mu_r)/\alpha}\left(\lambda (x-a)^{\alpha+\beta}\right).\n
 \eea






\newpage




\begin{figure}[ht] \centering
\includegraphics*[width=8cm,height=8cm]{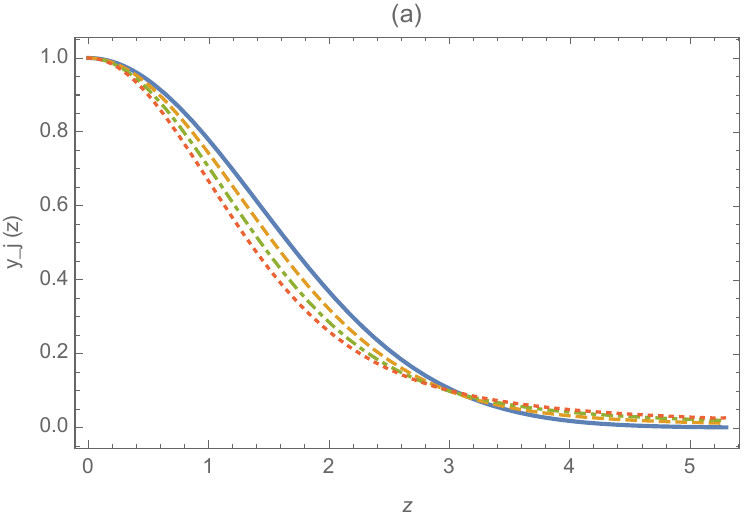}\hspace{1cm}
\includegraphics*[width=8cm,height=8cm]{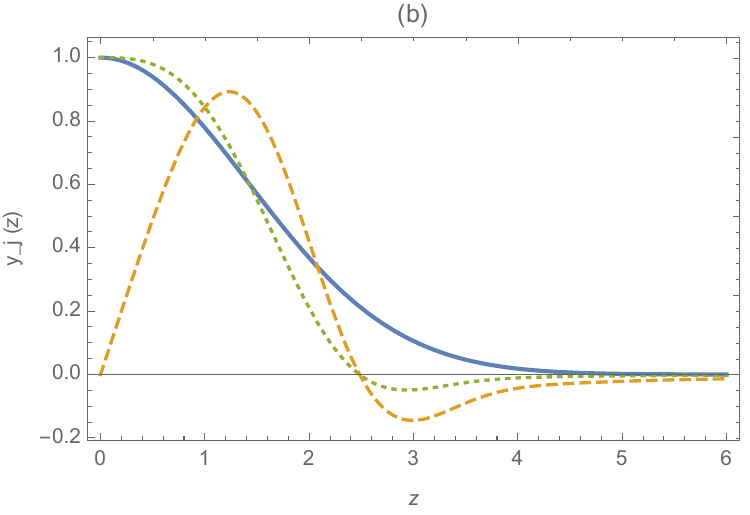}\\ 
\vskip 1cm
\includegraphics*[width=8cm,height=8cm]{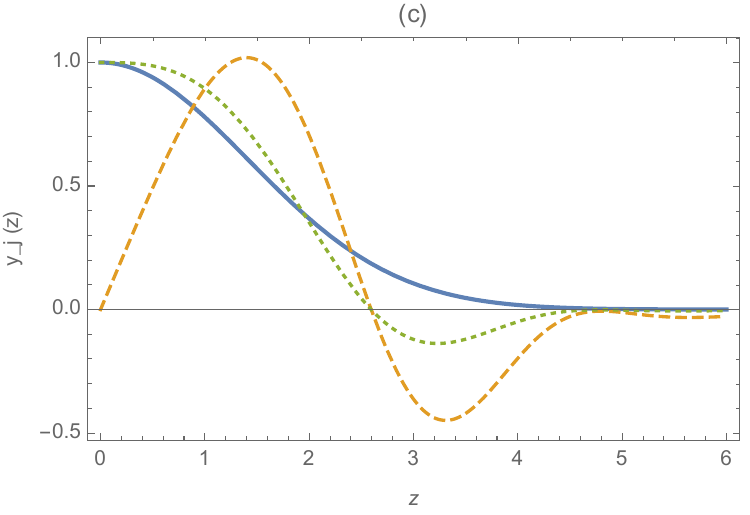}\hspace{1cm}
\includegraphics*[width=8cm,height=8cm]{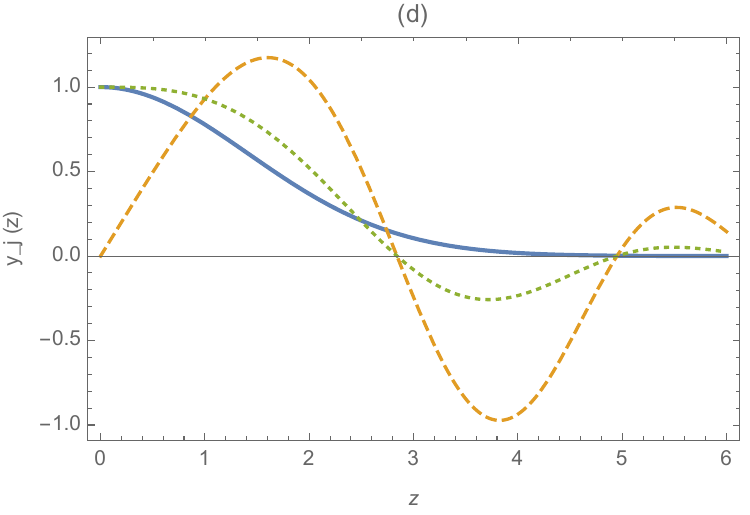}
\caption{Plots of $y_j(z)$ with $\alpha=1/\beta, q=n+1-\beta$ for different $\beta$ (solid line for $\beta=2, n=j=1$) : (a) $\beta=1.9$ (dashed), $1.8$ (dot-dashed), and $1.7$ (dotted).;  (b) $\beta=2.2$; (c) $\beta=2.4$, (d) $\beta=2.6$. For (b)-(d), $n=2$, $ j=1$ (dashed), and $2$ (dotted).
Number of terms used in the series expansion for the Kilbas-Saigo function is 400. }
\label{Fig1}
\end{figure}



\begin{figure}[ht] \centering
\includegraphics*[width=8cm,height=8cm]{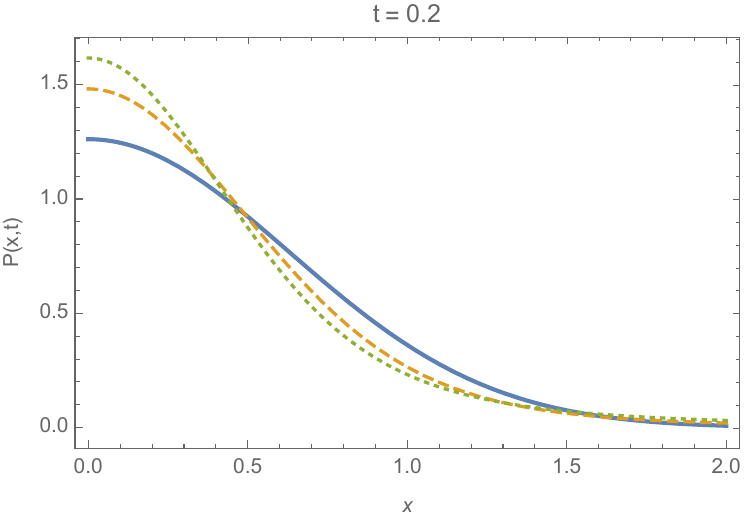}\hspace{1cm}
\includegraphics*[width=8cm,height=8cm]{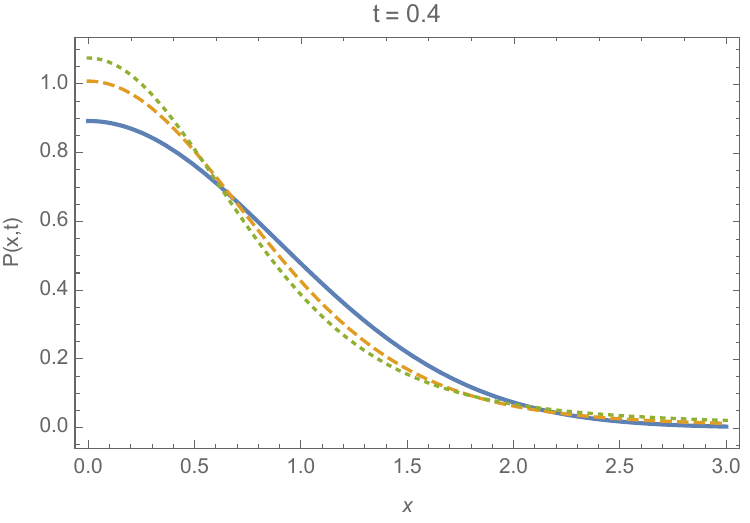}\\ 
\vskip 1cm
\includegraphics*[width=8cm,height=8cm]{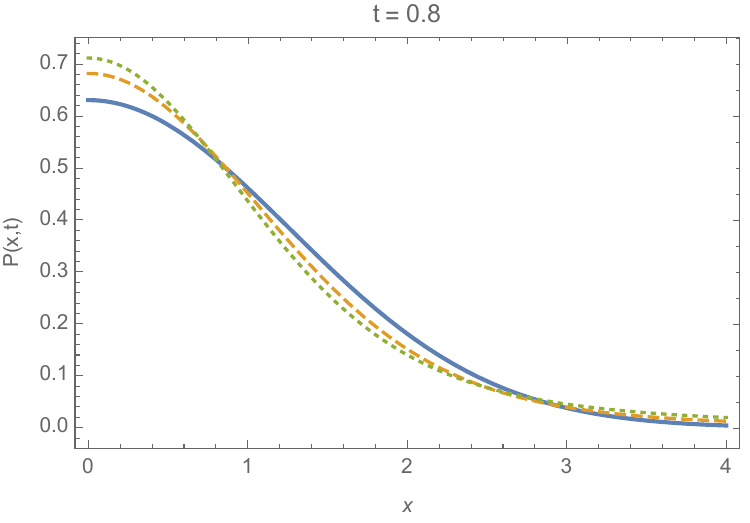}\hspace{1cm}
\includegraphics*[width=8cm,height=8cm]{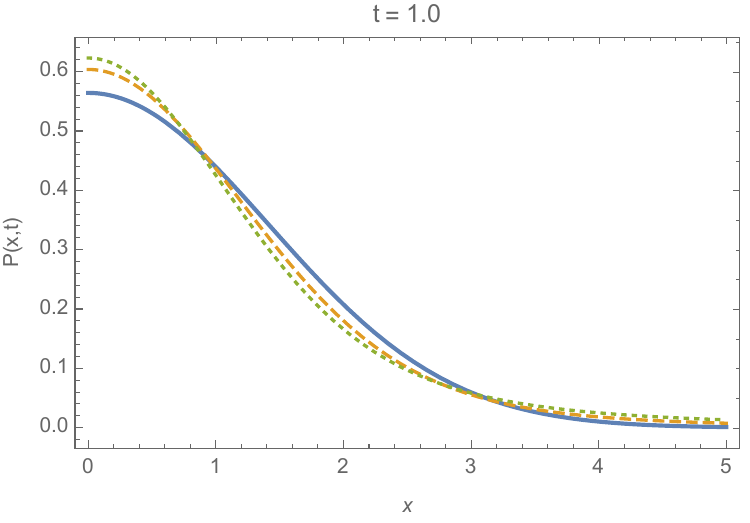}
\caption{Plots of $P(x,t)$ at different times for deformed Brownian-type diffusion, Eq.\,(\ref{P1}) with $q=2-\beta, \alpha = 1/\beta$, and $\beta= 2$  (solid), $1.8$ (dashed), and $1.7$ (dotted).  Number of terms used in the series expansion for the Kilbas-Saigo function is 400. }
\label{Fig2}
\end{figure}



\begin{figure}[ht] \centering
\includegraphics*[width=8cm,height=8cm]{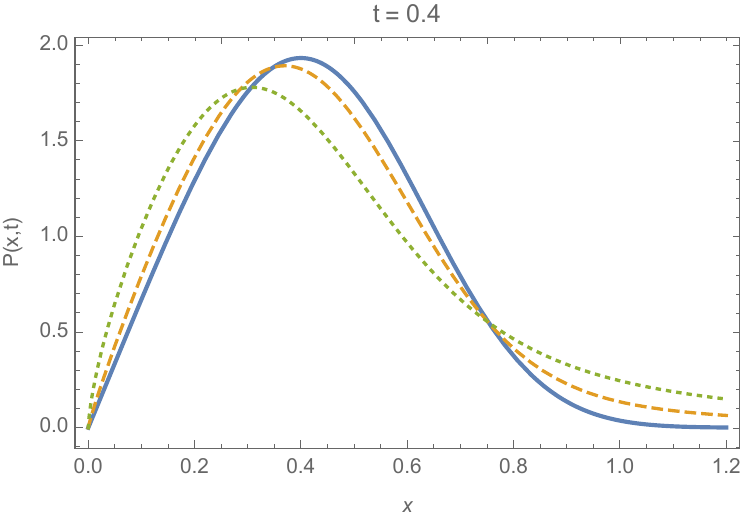}\hspace{1cm}
\includegraphics*[width=8cm,height=8cm]{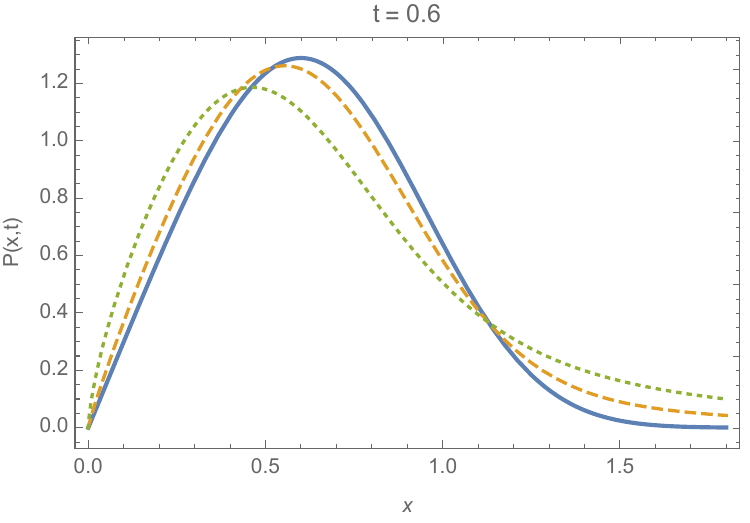}\\
\vskip 1cm
\includegraphics*[width=8cm,height=8cm]{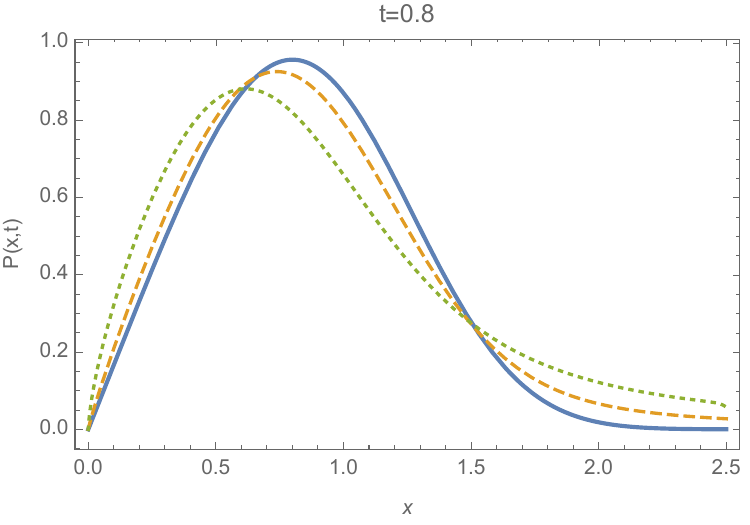}\hspace{1cm}
\includegraphics*[width=8cm,height=8cm]{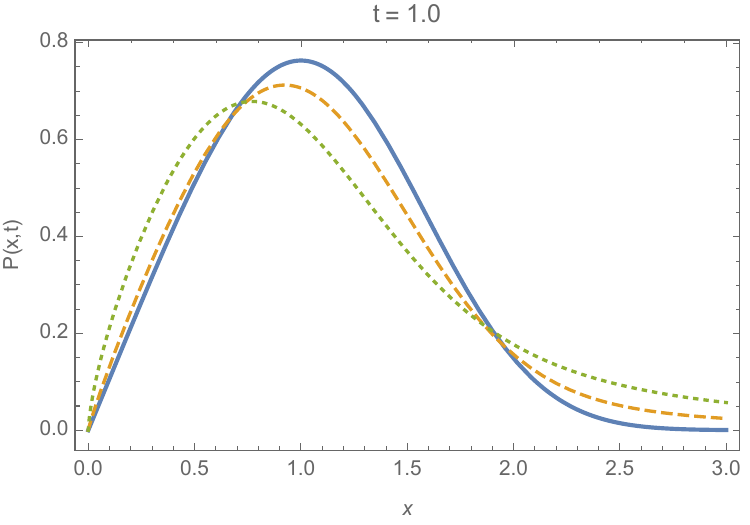}
\caption{Plots of $P(x,t)$  at different times for deformed non-Brownian type diffusion  Eq.\,(\ref{P1}) with $\alpha = 1, q=1\neq 2-\beta$, and $\beta= 2$  (solid), $1.9$ (dashed), and $1.7$ (dotted). Number of terms used in the series expansion for the Kilbas-Saigo function is 400. }
\label{Fig3}
\end{figure}



\begin{figure}[ht] \centering
\includegraphics*[width=8cm,height=8cm]{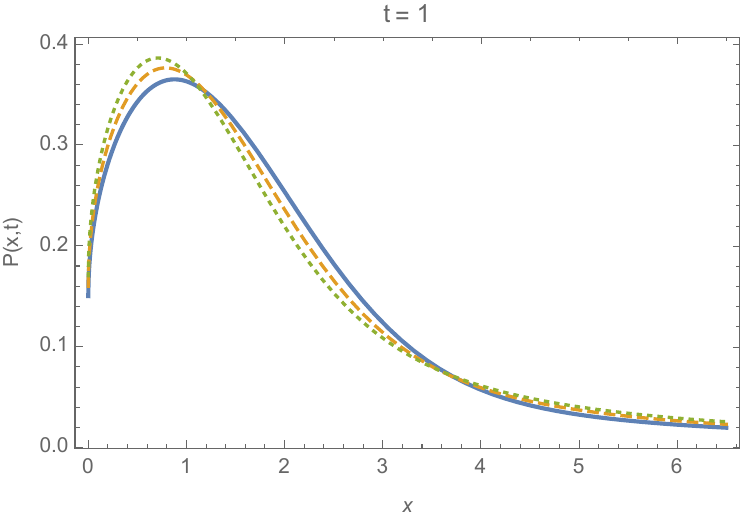}\hspace{1cm}
\includegraphics*[width=8cm,height=8cm]{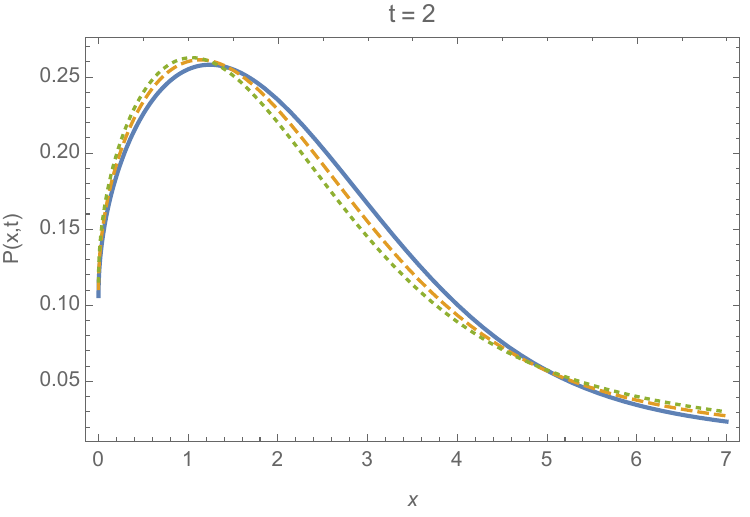}\\
\vskip 1cm
\includegraphics*[width=8cm,height=8cm]{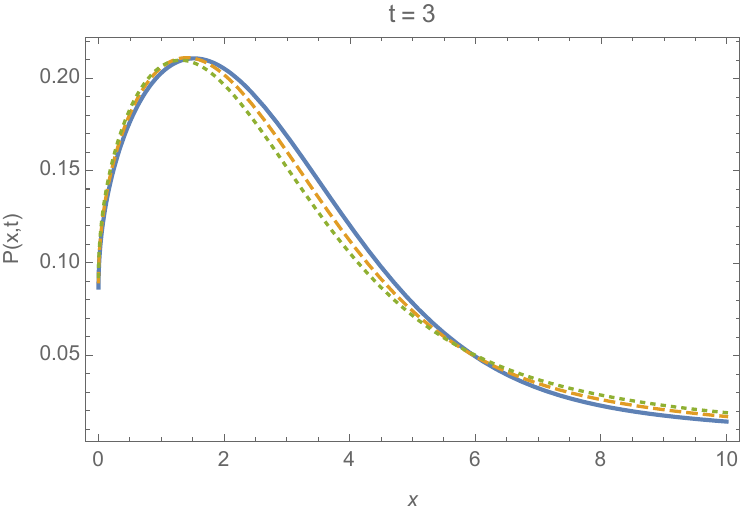}\hspace{1cm}
\includegraphics*[width=8cm,height=8cm]{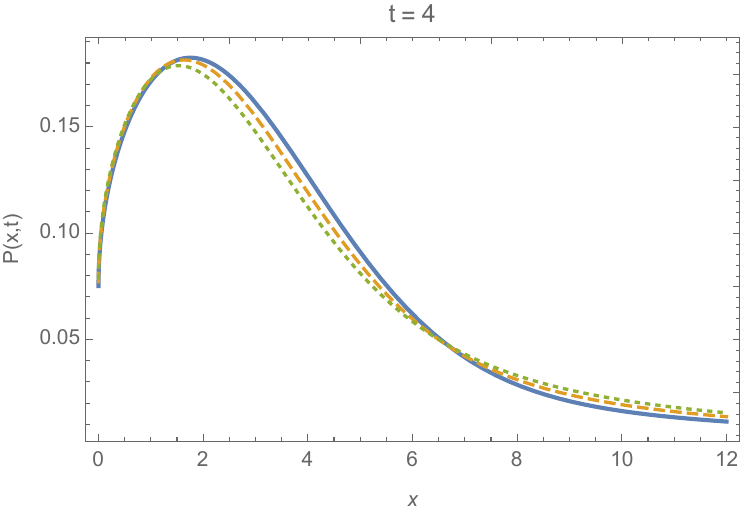}
\caption{Plots of $P(x,t)$  at different times for deformed non-Fokker-Planck type  reaction-diffusion Eq.\,(\ref{P2}) with $\mu_r=-0.5, \alpha = 1/\beta, q=2-\beta, c_1=c_2=1$, and $\beta= 2$  (solid), $1.9$ (dashed), and $1.8$ (dotted). Number of terms used in the series expansion for the Kilbas-Saigo function is 400. }
\label{Fig4}
\end{figure}

\end{document}